\documentclass[10pt,preprint,showpacs,preprintnumbers,amsmath,amssymb]{revtex4}
\usepackage{graphicx,epsfig}
\usepackage{subfigure}
\newcommand{\Mpc}{{\mathrm{Mpc}}}

\newcommand{\sca}{{\mathrm{S}}}
\newcommand{\ten}{{\mathrm{T}}}

\newcommand{\ex}{{\mathrm{ex}}}

\newcommand{\ji}{{\mathrm{j}}}

\begin{document}

\title{Computation of inflationary cosmological perturbations in the power-law
inflationary model using the phase-integral
method}

\author{Clara Rojas}
\email{clararoj@gmail.com}
\author{V\'ictor M. Villalba}
\email{villalba@ivic.ve}
\affiliation{Centro de F\'isica IVIC Apdo 21827, Caracas 1020A, Venezuela}

\date{\today}

\begin{abstract}
The phase-integral approximation devised by Fr\"oman and Fr\"oman,
is used for computing cosmological  perturbations in
the power-law inflationary model. The phase-integral formulas  for
the scalar and  tensor power spectra are explicitly obtained up to
ninth-order of the phase-integral approximation. We show that, the
phase-integral approximation exactly reproduces the shape of the
power spectra for scalar and tensor perturbations as well as the
spectral indices. We compare the accuracy of the phase-integral
approximation with the results for the power spectrum obtained with
the slow-roll and uniform approximation methods.
\end{abstract}


\maketitle
\section{Introduction}

The results reported by WMAP favor inflation \cite{spergel:2006}
over other cosmological scenarios.  The data is consistent with a
flat universe and with an almost scale invariant spectrum for the
primordial perturbations. The spectrum of the perturbations
generated during inflation depends on the model, therefore it is
important to predict the power spectrum of the cosmological
perturbations for a variety of inflationary scenarios. In general,
most of the inflationary models are not analytically solvable and
approximate or numerical methods are mandatory. Traditionally, the
method of approximation applied in inflationary cosmology is the
slow-roll approximation \cite{stewart:1993}. Recently, some authors
have applied semiclassical methods, such as the WKB method with the
Langer modification \cite{langer:1937,martin:2003A,casadio:2005A},
and the method of uniform approximation \cite{habib:2002,
habib:2004,habib:2005B}. In the present article we propose an
alternative method of approximation for the study of cosmological
perturbations during inflation, this method is based on the phase
integral approximation \cite{froman:1965,froman:1996,froman:2002}, which
has been successfully applied in different problems in quantum
mechanics \cite{froman:1965}, and in the study of quasinormal modes
in black hole physics \cite{Anderson,Kostas}.

The  Friedmann-Robertson-Walker line element for a spatially flat
universe can be written as
\begin{equation}
\label{met} ds^2 = -dt^2 + a^2(t)\left[dr^2+ r^2(d\theta^2 +
\sin^2\theta d\phi^2)\right],
\end{equation}
where $a$ is the scale factor. In order to study cosmological
perturbations we consider perturbations of the spatially-flat
Friedmann-Robertson-Walker universe (\ref{met}).

Using a gauge invariant treatment of linearized fluctuations in the
metric and field equations \cite{mukhanov:1992},we have that the
power spectra involves the computation of the two-point function
\begin{equation}
\langle
0|u(\eta,x),u(\eta,x+r)|0\rangle=\int_{0}^{\infty}\frac{dk}{k}\frac{sin
kr}{kr}P_{u}(\eta,k).
\end{equation}

The scalar  density perturbations are described by the function
$u_k=a\Phi/\phi'$, where $\Phi$ is a gauge-invariant variable
corresponding to the Newtonian potential, and  $\phi$ is the scalar
field. The equations of motion for the perturbation $u_k$ in a
universe dominated by a scalar field $\phi$ are given by

\begin{equation}\label{dotdotuk}
u_k''+\left(k^2-\frac{z_{S}''}{z_{S}}\right)u_k=0,
\end{equation}
where $z_{S}=a\phi'/\mathcal{H}$, $\mathcal{H}=a'/a$, and the prime
indicates derivative with respect to the conformal time $\eta=\int
dt/a$.

For the tensor perturbations (gravitational waves) we introduce a
function $v_k=ah$, where $h$ is the amplitude of the gravitational
wave. The tensor perturbations obey  a second order differential
equation analogous to Eq. (\ref{dotdotuk})

\begin{equation}\label{dotdotvk}
v_k''+\left(k^2-\frac{a''}{a}\right)v_k=0.
\end{equation}
Considering the limits  $k^2\gg|z_{S}''/z_{S}|$ (short wavelength) and
$k^2\ll|z_{S}''/z_{S}|$  (long wavelength), we have that  the  solutions to
Eq. (\ref{dotdotuk}) exhibit the following asymptotic behavior

\begin{equation}
\label{boundaryi}
 u_k\rightarrow \frac{e^{-ik\eta}}{\sqrt{2k}}
\quad \left(k^2\gg|z_{S}''/z_{S}|, -k\eta\rightarrow \infty \right),
\end{equation}

\begin{equation}
\label{boundary_0} u_k\rightarrow A_k z_{S}  \quad \left(k^2\ll|z_{S}''/z_{S}|,
-k\eta\rightarrow 0\right),
\end{equation}
the same asymptotic boundary conditions also hold for tensor
perturbations

Once the mode equations for scalar and tensor perturbations are
solved for different momenta $k$, the power spectra for scalar and
tensor modes are given by the expression
 \begin{eqnarray}
 \label{PS}
 P_S(k)&=& \lim_{-k\eta\rightarrow 0} \frac{k^3}{2 \pi^2}\left|\frac{u_k(\eta)}{z_{S}(\eta)} \right|^2,\\
 \label{PT}
 P_T(k)&=& \lim_{-k\eta\rightarrow 0} \frac{k^3}{2 \pi^2}\left|\frac{v_k(\eta)}{a(\eta)} \right|^2.
 \end{eqnarray}

The spectral indices are defined as \cite{liddle:2000}

\begin{equation}
\label{ns}
 n_S(k)=1+\frac{d \ln P_S(k)}{d \ln k},
\end{equation}
\begin{equation}
\label{nt}
 n_T(k)=\frac{d \ln P_T(k)}{d \ln k}.
\end{equation}
Running of the spectral indices is given by the second logarithmic
derivative of the power spectra

\begin{eqnarray}
 \alpha_S(k)=\frac{d \ln n_S(k)}{d \ln k},\\
 \alpha_T(k)=\frac{d \ln n_T(k)}{d \ln k}.
\end{eqnarray}
The power spectrum is usually fitted using the ansatz

\begin{equation}
P(k)=A_{fit}\left(\frac{k}{k_{*}}\right)^{n_{fit}+(1/2)\alpha_{fit}\ln(k/k_{*})},
\end{equation}
where the parameters $A_{fit}$, $n_{fit}$, and $\alpha_{fit}$ are
fitted with the observational data.  The spectral index is evaluated
at the value $k_{*}$ \cite{Lidsey}

\begin{equation}
n_{S}(k_{*})=1+n_{fit}.
\end{equation}
The running is parametrized by $\alpha_{fit}$. It is not possible to
have a constant $n$ and nonzero $\alpha$ given the definitions for
$P(k)$ and $n_{S}$. This inconsistency gives as a result a
uncontrolled growth of errors away from the parameter $k_{*}$.  It
is purpose of the present paper to show how to compute approximate
solutions for the scalar and tensor power spectra with the help of
the phase-integral approximation method. The article is structured
as follows: In section II we give an introductory review of  the
phase-integral method and the connection formulas. In Sec. III we
apply the phase-integral approximation to the power-law inflationary
model.  In Sec. IV we compare the results
for the power-spectra obtained using the phase-integral approach
with those computed with the slow-roll and uniform-approximation
methods.
\section{The phase-integral method}
Let us consider the differential equation

\begin{equation}\label{eq_ori}
\frac{d^2u_k}{dz^2}+R(z) u_k=0,
\end{equation}
where  $R(z)$ is an analytic function of  $z$. In order to obtain an
approximate solution to Eq. (\ref{eq_ori}), we are going to use the
phase integral method  developed by Fr\"oman
\cite{froman:1965,froman:1966B}. The phase integral approximation,
generated using a non specified base solution  $Q(z)$,  is a linear
combination of the phase integral functions
\cite{froman:1974A,froman:1996}, which exhibit the following  form

\begin{equation}\label{pi_uk}
 u_k=q^{-1/2}(z) \exp\left[\pm i\, \omega(z) \right],
 \end{equation}
where

 \begin{equation}\label{omega}
 \omega(z)=\int^z q(z) dz.
 \end{equation}
Substituting (\ref{pi_uk}) into (\ref{eq_ori}) we obtain that the
exact phase integrand $q(z)$ must be a solution of the differential
equation

\begin{equation}\label{q_ori}
q^{-3/2}(z)\frac{d^2}{dz^2}q^{-1/2}(z)+\frac{R(z)}{q^2(z)}-1=0.
\end{equation}
For any solution of Eq. (\ref{q_ori}) the functions (\ref{pi_uk}),
are linearly independent, the linear combination of the functions
$u_k$ represents a local solution. In order to solve the global
problem we choose a linear combination of phase integral solutions
representing the same solution in different regions of the complex
plane. This is known as the Stokes phenomenon \cite{froman:1965}.

If we have a function $Q(z)$ which is an approximate solution of Eq.
(\ref{q_ori}), the quantity $\epsilon_0$, obtained after
substituting $Q(z)$ into Eq. (\ref{q_ori})

\begin{equation}\label{epsilon_0}
\epsilon_0=Q^{-3/2}(z)\frac{d^2}{dz^2}Q^{-1/2}(z)+\frac{R(z)-Q^2(z)}{Q^2(z)},
\end{equation}
is small compared to unity. We take into account the relative small
size of $\epsilon_0$ by considering it proportional to $\lambda^2$,
where $\lambda$ is a small parameter. The parameter $\epsilon_0$ is
small when $Q(z)$ is proportional to $1/\lambda$ and $R(z)-Q^2(z)$
is independent of $\lambda$, i.e. if  $R(z)$ is replaced by
$Q^2(z)/\lambda^2+\left[R(z)-Q^2(z)\right]$ in Eq. (\ref{eq_ori}).
Therefore, instead of considering Eq. (\ref{eq_ori}), we deal with
the auxiliary differential equation

\begin{equation}\label{eq_aux}
\frac{d^2u_k}{dz^2}+\left\{\frac{Q^2(z)}{\lambda^2}+\left[R(z)-Q^2(z)\right]\right\}
u_k=0,
\end{equation}
which reduces to Eq. (\ref{eq_ori}) when  $\lambda=1$.

Inserting the solutions (\ref{pi_uk}) into the auxiliary
differential equation (\ref{eq_aux}), we obtain the following
equation for $q(z)$

\begin{equation}
q^{1/2}\frac{d^2}{dz^2}q^{-1/2}-q^2+\frac{Q^2(z)}{\lambda^2}+R(z)-Q^2(z)=0,
\end{equation}
which is called the auxiliary $q$ equation.  After introducing the
new variable $\xi$,
\begin{equation}
\xi=\int^zQ(z)dz,
\end{equation}
we obtain
\begin{equation}\label{qQdif}
1-\left[\frac{q\lambda}{Q(z)}\right]^2+\epsilon_0\lambda^2+\left[\frac{q\lambda}{Q(z)}\right]^{1/2}\frac{d^2}{d\xi^2}\left[\frac{q\lambda}{Q(z)}\right]^{-1/2}\lambda^2=0,
\end{equation}
where $\epsilon_0$ is defined by Eq. (\ref{epsilon_0}). A formal
solution of Eq. (\ref{qQdif}) is obtained after the identification

\begin{equation}\label{qlambdaQ}
\frac{q\lambda}{Q}=\sum^\infty_{n=0} Y_{2n}\lambda^{2n}.
\end{equation}
Substituting Eq. (\ref{qlambdaQ}) into Eq. (\ref{qQdif}), we obtain

\begin{equation}\label{recurrence}
1-\left(\sum_n
Y_{2n}\lambda^{2n}\right)^2+\epsilon_0\lambda^2+\left(\sum_n
Y_{2n}\lambda^{2n}\right)^{1/2}\frac{d^2}{d\xi^2}\left(\sum_n
Y_{2n}\lambda^{2n}\right)^{-1/2}=0.
\end{equation}

Using computer manipulation algebra it is straightforward to obtain
the coefficients $Y_{2n}$. The first values are
\cite{froman:1966B,campbell:1972}

\begin{eqnarray}
\label{Y0}
Y_0&=&1,\\
\label{Y2}
Y_2&=&\frac 1 2 \epsilon_0,\\
Y_4&=&-\frac 1 8 \left(\epsilon_0^2+\epsilon_2 \right),\\
Y_6&=&\frac 1 {32} \left(2 \epsilon_0^2 + 6 \epsilon_0\epsilon_2+5\epsilon_1^2+\epsilon_4\right),\\
\label{Y8} Y_8&=&-\frac 1 {128} \left(5 \epsilon_0^4+30
\epsilon_0^2\epsilon_2+50\epsilon_0\epsilon_1^2+10\epsilon_0\epsilon_4+28\epsilon_1\epsilon_3+19
\epsilon_2^2+\epsilon_6\right),
\end{eqnarray}
where  $\epsilon_\nu$ is defined as

\begin{equation}\label{epsilon_nu}
\epsilon_\nu=\frac{1}{Q(z)}\frac{d\epsilon_{\nu-1}}{dz}, \quad \nu
\ge 1.
\end{equation}
Truncating the series (\ref{qlambdaQ}) at $n=N$ with $\lambda=1$ we
obtain

\begin{equation}\label{qQ}
q(z)=Q(z)\sum^N_{n=0} Y_{2n},
\end{equation}
Substituting (\ref{qQ}) in (\ref{omega}) we have that

\begin{equation}\label{omegadef}
\omega(z)=\sum_{n=0}^N \omega_{2n}(z),
\end{equation}
where

\begin{equation}\label{omega_sum}
 \omega_{2n}(z)=\int^z Y_{2n}Q(z) dz.
\end{equation}

From  (\ref{qQ}), (\ref{omegadef}) and (\ref{pi_uk}) we obtain a
phase integral approximation of order $2N+1$ generated with the help
of the base function $Q(z)$.

The base function $Q(z)$ is not specified and its selection depends
on the problem  in question. In many cases, it is enough to choose
$Q^2(z)=R(z)$, and the first-order phase integral approximation
reduces to the WKB approximation. In the first-order approximation
it is convenient to choose a root of $Q^2(z)$ as the lower
integration limit in expression (\ref{omega_sum}). However,  for
higher orders, i.e. for $2N+1>1$, this is not possible because the
function $q(z)$ is singular at the zeros of $Q^2(z)$. In this case,
it is convenient to express $\omega_{2n}(z)$ as a contour integral
over a two-sheet Riemann surface where $q(z)$ is single valued
\cite{froman:1966B}. We define

\begin{equation}
\omega_{2n}(z)=\frac{1}{2} \int_{\Gamma_{t}}Y_{2n}(z)Q(z)dz,
\end{equation}
where  $t$ is a zero of $Q^2(z)$ and  $\Gamma_{t}$ is an integration
contour  starting  at the point corresponding to $z$ over a Riemann
sheet adjacent to the complex plane, and that encloses the point
$t$, in the positive or negative sense and ends at the point $z$.

If the function $Q(z)$ is chosen conveniently, the quantity  $\mu$
defined by

\begin{equation}\label{mu}
\mu=\mu(z,z_0)=\left|\int_{z_0}^z\left|\epsilon(z)q(z)
dz\right|\right|,
\end{equation}
is much smaller than 1.  The function  $\epsilon(z)$ is given by the
left side of Eq. (\ref{q_ori})

\begin{equation}
\epsilon(z)=q^{-3/2}(z)\frac{d^2}{dz^2}q^{-1/2}(z)+\frac{R(z)}{q^2(z)}-1,
\end{equation}
where the integral $\mu$ measures  the accuracy of the
phase-integral approximation \cite{froman:2002}.

We assume that the function $Q^2(z)$ is real over the real axis.
Taking into account this restriction, we shall call turning point,
the zero of $Q^2(z)$. We want to know the connection formulas at
both sides of an isolated turning point $z_{ret}$, i.e., a turning
point which is located far from other turning points. We will adopt
the terms  ``classically permitted region'' and  ``classically
forbidden region'' in order to denote those ranges over the real
axis where $Q^2(z)>0$ and $Q^2(z)<0$, respectively.

The connection formula for an approximate solution that crosses the
turning point $z_{ret}$ from a classically permitted region to a
classically forbidden region is \cite{froman:1970A}

\begin{equation}\label{allowed-forbbiden}
\left|
q^{-1/2}(z)\right|\cos\left(\left|\omega(z)\right|+\frac{\pi}{4}\right)\rightarrow
\left| q^{-1/2}(z)\right|\exp\left[\left|\omega(z)\right|\right].
\end{equation}

The connection formula for an approximate solution that crosses the
turning point $z_{ret}$ from a classically forbidden region to a
classically permitted region is \cite{froman:1970A}

\begin{equation}\label{forbbiden-allowed}
\left| q^{-1/2}(z)\right|\exp\left[-\left|\omega(z)\right|\right]
\rightarrow 2 \left|
q^{-1/2}(z)\right|\cos\left(\left|\omega(z)\right|-\frac{\pi}{4}\right).
\end{equation}

It is important to emphasize the one-directional character of the
connection formulas (\ref{allowed-forbbiden}) and
(\ref{forbbiden-allowed}), this means that  the trace of the
solution should be done in the direction indicated by the arrows in
Eq. (\ref{allowed-forbbiden}) and Eq. (\ref{forbbiden-allowed}).

\subsection{The phase-integral power spectra}

We proceed to apply the phase integral approximation to Eq. (\ref{dotdotuk}) and Eq. (\ref{dotdotvk}). Since the
the square of the base function  $Q_{S,T}^2(k,\eta)$ possesses only  one turning point   $\eta_{_{S,T}}$
that can be obtained after solving the equation $Q_{S,T}^2(k,\eta)=0$, we can divide the axis $\eta$ into
two regions:

\begin{eqnarray}
\quad \eta_{_{S,T}}<-\eta<0&Q_{S,T}^2<0,\\
\quad-\eta<\eta_{_{S,T}},&Q_{S,T}^2>0.
\end{eqnarray}

The corresponding phase-integral solutions are:

For $\eta_{_{S,T}}<-\eta<0$

\begin{eqnarray}
\label{uk_right}
u_k^{phi}(k,\eta)&=&\frac{c_1}{2} \left|q_S^{-1/2}(k,\eta)\right|\exp\left(-\left|w_S(k,\eta)\right|\right)\\
\nonumber
&+&c_2 \left|q_S^{-1/2}(k,\eta)\right|\exp\left(\left|w_S(k,\eta)\right|\right),\\
\label{vk_right}
v_k^{phi}(k,\eta)&=&\frac{d_1}{2} \left|q_T^{-1/2}(k,\eta)\right|\exp\left(-\left|w_T(k,\eta)\right|\right)\\
\nonumber
&+&d_2 \left|q_T^{-1/2}(k,\eta)\right|\exp\left(\left|w_T(k,\eta)\right|\right).
\end{eqnarray}

For $-\eta<\eta_{_{S,T}}$

\begin{eqnarray}
\label{uk_left}
u_k^{phi}(k,\eta)&=&c_1 \left|q_S^{-1/2}(k,\eta)\right|\cos\left(\left|w_S(k,\eta)\right|-\frac{\pi}{4}\right)\\
\nonumber
&+&c_2 \left|q_S^{-1/2}(k,\eta)\right|\cos\left(\left|w_S(k,\eta)\right|+\frac{\pi}{4}\right),\\
\label{vk_left}
v_k^{phi}(k,\eta)&=&d_1 \left|q_T^{-1/2}(k,\eta)\right|\cos\left(\left|w_T(k,\eta)\right|-\frac{\pi}{4}\right)\\
\nonumber
&+&d_2 \left|q_T^{-1/2}(k,\eta)\right|\cos\left(\left|w_T(k,\eta)\right|+\frac{\pi}{4}\right).
\end{eqnarray}
where $c_1$, $c_2$ and $d_1$, $d_2$ are obtained after comparing the
asymptotic behavior of Eq.  (\ref{uk_left}) and Eq. (\ref{vk_left})
with  the asymptotic limit  given by Eq. (\ref{boundaryi}). In order
to calculate the power spectra we substitute the growing part of the
solutions (\ref{uk_right}) and (\ref{vk_right}) into  Eq. (\ref{PS})
and Eq. (\ref{PT}).  We obtain the following expressions

\begin{eqnarray}
\label{PS_pi}
P_S^{phi}(k)&=&\lim_{-k\eta\rightarrow 0} \frac{k^3}{2\pi^2} \frac{\left|c_2\right|^2}{\left|z_S(\eta)\right|^2}\left|q_S^{-1}(k,\eta)\right|\exp\left(2 \left|w_S(k,\eta) \right|\right),\\
\label{PT_pi}
P_T^{phi}(k)&=&\lim_{-k\eta\rightarrow 0} \frac{k^3}{2\pi^2} \frac{\left|d_2\right|^2}{\left|a(\eta)\right|^2}\left|q_T^{-1}(k,\eta)\right|\exp\left(2 \left|w_T(k,\eta) \right|\right).
\end{eqnarray}
The spectral indices $n_S$ and $n_T$ in the phase-integral
approximation can be obtained respectively from Eq.(\ref{ns}) and
(\ref{nt}) and are given by
\begin{eqnarray}
\label{nS_pi}
n_S^{phi}(k)&=&4+\lim_{-k\eta\rightarrow 0} \left[\frac{d\ln\left|q_S^{-1}(k,\eta)\right|}{d\ln k}+2\frac{d \left|w_S(k,\eta)\right|}{d\ln k}\right],\\
\label{nT_pi}
n_T^{phi}(k)&=&3+\lim_{-k\eta\rightarrow 0} \left[\frac{d\ln\left|q_T^{-1}(k,\eta)\right|}{d\ln k}+2\frac{d \left|w_T(k,\eta)\right|}{d\ln k}\right],
\end{eqnarray}
\section{Application to power-law inflation}\label{pipl}

The power-law inflationary model is a very simple model that allows
one to solve the horizon and flatness problem. Since this model does
not have a natural way of terminating the inflationary epoch, it is 
not physically acceptable, nevertheless its advantage lies in
the possibility of analytically computing the solutions to the
perturbation equations and the corresponding power spectra
\cite{abbott:1984,lucchin:1985}. The power-law model allows testing
approximations that are necessary in other inflationary models that
do not exhibit analytic solutions. In this model, the scale factor
is given by

\begin{equation}\label{a}
a(\eta)= l_0  \eta ^{\frac{1}{2}-\nu},
 \end{equation}
where $\nu=\frac{3}{2}+\frac{1}{p-1}$. We have to impose the
condition $p>1$ in order that Eq. (\ref{a}) satisfies the
inflationary condition  $\ddot{a}>0$.

Using the power-law scale factor (\ref{a}) we find that
$z_{S}=l_0M_{Pl}\sqrt{\frac{2}{p}}\eta^{\frac{1}{2}-\nu}$. Since, for
this model, the differential equations governing the scalar and
tensor perturbations are identical, we make the identification
$u_k=v_k$ with
\begin{equation}\label{uk}
\frac{d^2u_k}{d\eta^2}+\left[k^2-\frac{\left(\nu^2-\frac{1}{4}\right)}{\eta^2}\right]
u_k=0,
\end{equation}
where the function $u_k$ in Eq. (\ref{uk}) satisfies the boundary
conditions (\ref{boundaryi}) and (\ref{boundary_0}).

Eq. (\ref{uk}) can be exactly solved. The exact solution, satisfying
the boundary conditions (\ref{boundaryi}) and (\ref{boundary_0}) can
be expressed in terms of a fractionary Hankel function
\cite{liddle:2000}

\begin{equation}
u_k^{ex}(\eta)=\frac{\sqrt{\pi}}{2}
\exp\left[i\left(\nu+\frac{1}{2}\right)\frac{\pi}{2}\right]
\sqrt{-\eta} H_{\nu}^{(1)}(-k \eta).
\end{equation}
The exact power energy spectra are given by
\begin{eqnarray}
P_S^{ex}(k)&=&\frac{1}{l_0^2 M_{Pl}^2} \mathrm{g^{ex}}(\nu) k^{3-2\nu},\\
P_T^{ex}(k)&=&\frac{1}{l_0^2} h^{ex}(\nu)  k^{3-2\nu},
\end{eqnarray}
where

\begin{eqnarray}
g^{ex}(\nu) &=& \left(\frac{1-2\nu}{3-2\nu}\right)\left[\frac{2^{\nu-2}\Gamma(\nu)}{2\pi\Gamma(\frac{3}{2})}\right]^2,\\
h^{ex}(\nu)&=
&\left[\frac{2^{\nu-3/2}\Gamma(\nu)}{2\pi\Gamma(\frac{3}{2})}\right]^2,
\end{eqnarray}
the corresponding spectral indices are

\begin{equation}
n_S^{ex}(k)=3-\frac{2p}{p-1},
\label{indices1}
\end{equation}
\begin{equation}
n_T^{ex}(k)=2-\frac{2p}{p-1}.
\label{indices2}
\end{equation}
In order to apply the phase-integral method  it is useful to
introduce the variable $z=k\eta$. The function $R(z)$ has the form

\begin{equation}\label{R_z}
R(z)=a_0+\frac{a_{-2}}{z^2},
\end{equation}
where $a_0=1$ and $a_{-2}=\frac{1}{4}-\nu^2$ are constants. In order
to solve  Eq. (\ref{uk}) with the help of the phase integral
approximation we need to choose the base function  $Q(z)$. If we
choose the square of the base function as  $Q^2(z)=R(z)$ one obtains
that the  quantity $\mu(z,z_{0})$ given by (\ref{mu}) is singular at the
origin, which is the place where the boundary condition
(\ref{boundary_0}) has to be imposed. We can circumvent this problem
making the following choice for the square of $Q(z)$
\cite{froman:2002}

\begin{equation}\label{Q^2_z}
Q^2(z)=b_0+\frac{b_{-2}}{z^2},
\end{equation}
where  $b_0$ and  $b_{-2}$ are constants. The coefficients in Eq.
(\ref{Q^2_z}) are chosen in a way that makes the phase integral
approximation valid to any order as  $-z\rightarrow 0$. To
verify the validity of the approximation at any order in a vicinity
of zero, we require that the integral $\mu(z,z_0)$ defined by Eq.
(\ref{mu}) be finite as  $-z\rightarrow 0$. For the problem we are
discussing, it is enough to verify that the  integral $\mu(z,z_0)$
particularized to first order be valid in a vicinity of the origin.
This condition guarantees that, at any order of approximation, the
integral $\mu(z,z_0)$ be defined in the vicinity of the origin. To
first order, we have that $q(z)=Q(z)$, and
$\epsilon=\epsilon_0$, then $\mu(z,z_0)$ can be written as

\begin{equation}
\label{mu_first} \mu(z,z_0)=\left|\int_{z_0}^{z} \left|
\epsilon_0(z) Q(z) dz\right|\right|.
\end{equation}
Making the corresponding computations we obtain

\begin{equation}
\epsilon_0=\frac{4 a_{-2} -4 b_{-2}-1}{4b_{-2}} +
\left[\frac{\left(4b_{-2}-4a_{-2}-3\right)b_0}{4
b_{-2}^2}+\frac{a_0-b_0}{b_{-2}} \right]z^2+{\mathcal{O}}(z^4),
\end{equation}
with
\begin{equation}
\label{epsilon0Q} \epsilon_0 Q=\frac{4 a_{-2} -4
b_{-2}-1}{4b_{-2}^{1/2}} z^{-1} +
\left[\frac{\left(4b_{-2}-4a_{-2}-7\right)b_0}{8
b_{-2}^{3/2}}+\frac{a_0-b_0}{b_{-2}^{1/2}}
\right]z+{\mathcal{O}}(z^3).
\end{equation}
With the help of Eq. (\ref{epsilon0Q}) and the expression
(\ref{mu_first}) we get that  the condition of validity of the first
order approximation, as $-z \rightarrow 0$, implies that

\begin{equation}
\label{b-2} b_{-2}=a_{-2}-\frac{1}{4}.
\end{equation}
Using Eq. (\ref{R_z})-(\ref{Q^2_z}) and Eq. (\ref{b-2}) we have that
\begin{equation}
\displaystyle\lim_{-z\rightarrow 0}
z^2\left[Q^2(z)-R(z)\right]=-\frac{1}{4}.
\end{equation}

The following election for  $Q^2(z)$ is  convergent as  $-z
\rightarrow 0$, and it is valid to any order of approximation

\begin{equation}
\label{Q2}
Q^2(z)=R(z)-\frac{1}{4 z^2}.
\end{equation}
The equation governing the modes  $k$ for the scalar and tensor is
\begin{equation}
\label{pl_ddotuk_pi} \frac{d^2 u_k}{d z^2}+Q^2(z) u_k=0,
\end{equation}
where

\begin{equation}
\label{pl_Q^2_z} Q^2(z)=1-\frac{\nu^2}{z^2},
\end{equation}
therefore, the phase-integral approximation is valid as $-z
\rightarrow 0$, where the boundary condition (\ref{boundary_0})
should be imposed.

\begin{figure}[htbp]
\begin{center}
\subfigure[]{\includegraphics[scale=0.3]{pl_Q2.eps}}\\
\subfigure[]{\includegraphics[scale=0.3]{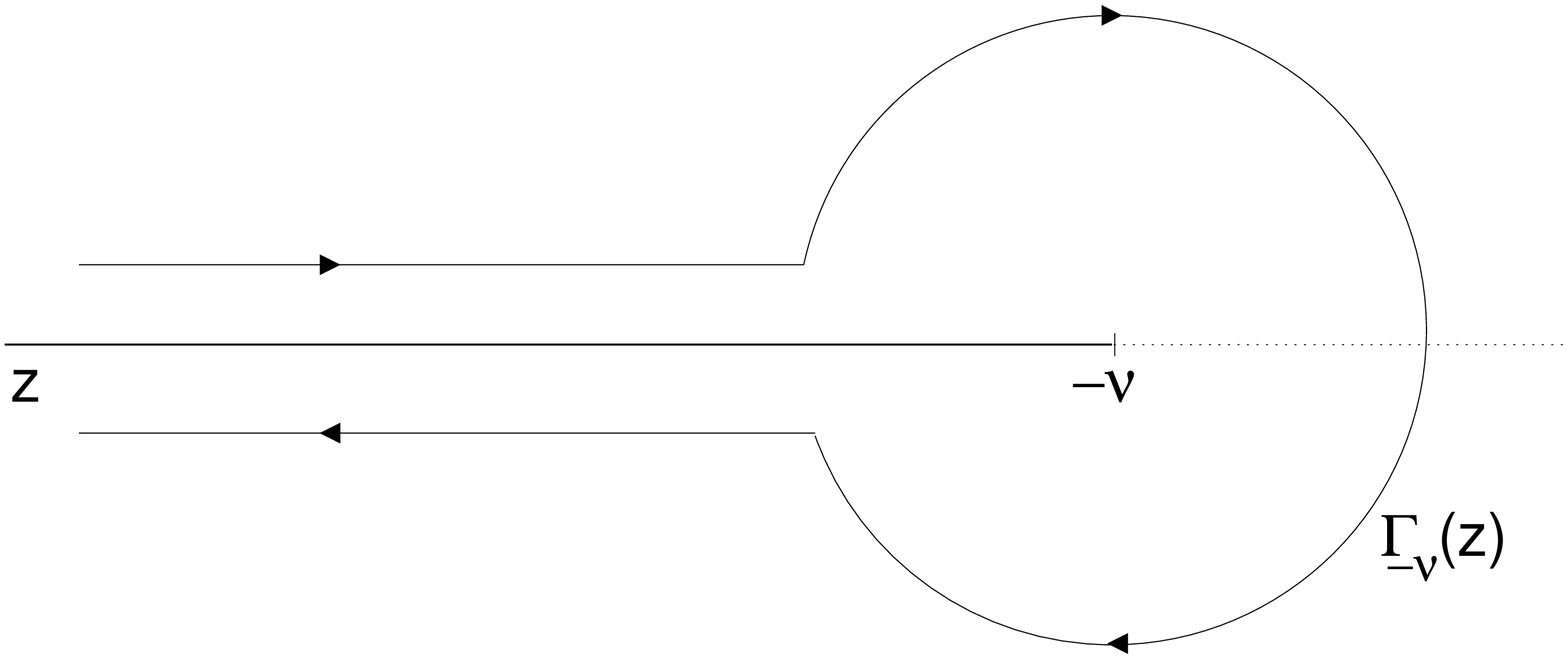}}
\subfigure[]{\includegraphics[scale=0.3]{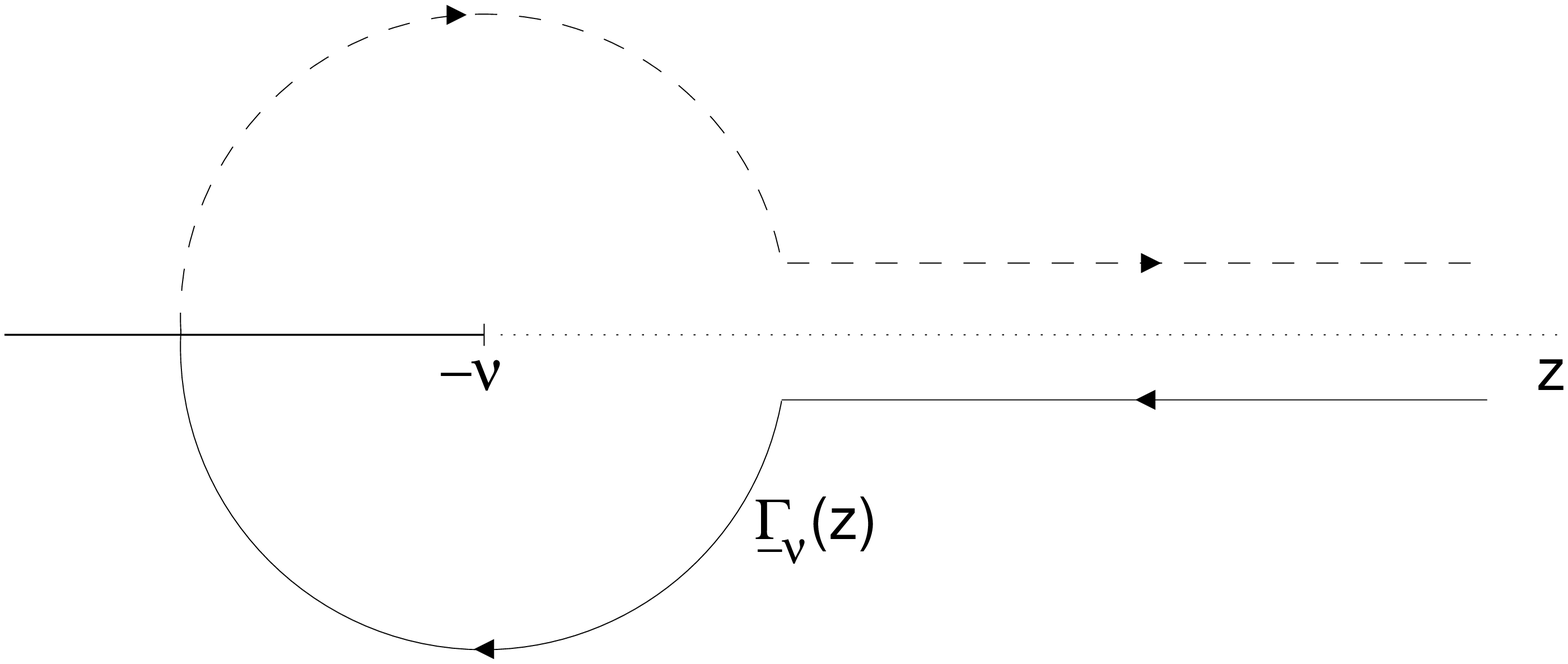}}
\caption{(a) Behavior of the  function $Q^2(z)$ for $\nu<0$ (b)
Contour of integration $\Gamma_{-\nu}(z)$ for $-z<-\nu$. (c) Contour
of integration $\Gamma_{-\nu}(z)$ for $-\nu<-z<0$. The dashed line
indicates  the path of integration on the second Riemann sheet.}
\label{fig:Q^2}
\end{center}
\end{figure}

The square of the base function $Q^2(z)$  exhibits two turning
points $z_{ret}=\pm\nu$. Since we are interested in the limit
$-z\rightarrow 0$, we choose to work with the negative turning
point. This turning point corresponds to the horizon $k=a H$
($z=-\nu$). The solution is defined in two ranges: On the left of
the turning point, corresponding to scales lower than the horizon,
we have the classically permitted region $Q^2(z)>0$ where the
solution oscillates. On the right of the turning point $-\nu<-z<0$,
corresponding to scales larger than the horizon, we have the
classically forbidden region $Q^2(z)<0$, where the solution grows or
decays exponentially. Fig. \ref{fig:Q^2} (a) shows the two ranges
where the solution is defined. In the phase-integral approximation,
for $-\nu<-z<0$,  the solutions associated with the modes $k$
for the scalar and tensor
perturbations are given by  (\ref{uk_right}) and (\ref{vk_right}) respectively. For
$-z < -\nu$, the solutions are  (\ref{uk_left}) and   (\ref{vk_left}).

\bigskip
Using Eq. (\ref{qQ}) we have that the ninth-order approximation
($2N+1=9\rightarrow N=4$) of the function $q(z)$ has the form

\begin{equation}\label{q_exp}
q(z)=\sum_{n=0}^4 Y_{2n} Q(z)=\left(Y_0+Y_2+Y_4+Y_6+Y_8\right) Q(z).
\end{equation}

In order to compute $q(z)$ in Eq. (\ref{q_exp}), we need to compute
the coefficients $Y_{2n}$ (\ref{Y0})-(\ref{Y8}). The calculation of
$Y_{2n}$  requires the knowledge of the coefficients $\epsilon_\nu$.
Using Eq. (\ref{epsilon_0}) and Eq. (\ref{epsilon_nu}) we obtain

\begin{align}
\label{epsilon0}
\epsilon_0&=\frac{z^2}{\left(z^2-\nu^2\right)^3}\left(\frac{z^2}{4}+\nu^2\right),\\
\epsilon_1&=-\frac{z^2}{\left(z^2-\nu^2\right)^{\frac{9}{2}}}\left(\frac{z^4}{2}+5\nu^2 z^2+2\nu^4\right),\\
\epsilon_2&=\frac{z^2}{\left(z^2-\nu^2\right)^{6}}\left(\frac{3 z^6}{2}+28\nu^2z^4+34\nu^4z^2+4\nu^6\right),\\
\epsilon_3& =-\frac{z^2}{\left(z^2-\nu^2\right)^{\frac{15}{2}}} \left(6z^8+180\nu^2z^6+440\nu^4z^4 +176\nu^6z^2+ 8\nu^8\right),\\
\epsilon_4&=\frac{z^2}{\left(z^2-\nu^2\right)^9}\left(30z^{10}+1320\nu^2z^8+5400\nu^4z^6+4576\nu^6 z^4+808\nu^8z^2+16\nu^{10}\right),\\
\epsilon_5&=-\frac{z^2}{\left(z^2-\nu^2\right)^{\frac{21}{2}}}\left(180z^{12}+10920\nu^2z^{10}+67200\nu^4z^8+98112\nu^6z^6+38768\nu^8z^4+\right.\\
\nonumber&\left.+3488\nu^{10}z^2+32\nu^{12}\right),\\
\label{epsilon6}
\epsilon_6&=\frac{z^2}{\left(z^2-\nu^2\right)^{12}}\left(1260z^{14}+100800\nu^2z^{12}+870240\nu^4z^{10}+1947456\nu^6z^8+\right.\\
\nonumber&+\left.1366416\nu^8z^6+291904\nu^{10}z^4+14560\nu^{12}z^2+64\nu^{14}\right).
\end{align}

Inserting Eq (\ref{epsilon0})-(\ref{epsilon6}) into Eq.
(\ref{Y2})-(\ref{Y8}) we obtain that the first four
$Y_{2n}$ are

\begin{align}
Y_2&=\frac{z^2}{\left(z^2-\nu^2\right)^3}\left(\frac{z^2}{8}+\frac{\nu^2}{2}\right),\\
Y_4&=-\frac{z^2}{\left(z^2-\nu^2\right)^6}\left(\frac{25z^6}{128}+\frac{57\nu^2z^4}{16}+\frac{35\nu^4z^2}{8}+\frac{\nu^6}{2}\right),\\
Y_6&=\frac{z^2}{\left(z^2-\nu^2\right)^9}\left(\frac{1073z^{10}}{1024}+\frac{11171\nu^2z^8}{256}+\frac{11511\nu^4z^6}{64}+\frac{611\nu^6z^4}{4}+\frac{213\nu^8z^2}{8}+\frac{\nu^{10}}{2}\right),\\
Y_8&=-\frac{z^2}{\left(z^2-\nu^2\right)^{12}}\left(\frac{375733z^{14}}{32768}+\frac{1752101\nu^2z^{12}}{2048}+\frac{7572705\nu^4z^{10}}{1024}+\frac{4252813\nu^6z^8}{256}+\right.\\\nonumber
&\left.+\frac{1491943\nu^8z^6}{128}+\frac{39561\nu^{10}z^4}{16}+\frac{967\nu^{12}z^2}{8}+\frac{\nu^{14}}{2}\right).
\end{align}

After computing  the coefficients $Y_{2n}$ up to $N=4$ we obtain a
ninth-order approximation for $q(z)$. The next step is to compute
$\omega(z)$. Fig. \ref{fig:Q^2} (b)-(c) show the contours of
integration used for computing the integral $\omega(z)$ beginning
with the second-order of approximation. Thus we have

\begin{eqnarray}
\omega(z)&=&w_0(z)+\sum_{n=1}^{4} w_{2n}(z),\\
&=&\int_{-\nu}^{-z}Q(z)dz+\frac{1}{2}\sum_{n=1}^4\int_{\Gamma_{-\nu}}Y_{2n}Q(z)dz.
\end{eqnarray}

The expressions for $\omega_{2n}(z)$, written  in the variable
$\eta$, up to $N=4$, are:

\begin{align}
\omega_0&= \sqrt{(-k\eta)^2-\nu^2}- \nu \arccos\frac{\nu}{-k\eta},\quad -k\eta<-\nu, \\
i\omega_0&=-\sqrt{\nu^2-(-k\eta)^2}+\nu\ln\left|\frac{\nu+\sqrt{\nu^2-(-k\eta)^2}}{-k\eta}\right|,\quad -\nu<-k\eta<0,\\
\label{omega_2} \left.\begin{array}{cc}
\omega_2\\
i\omega_2&
\end{array}\right\}&=\mp \frac{1}{24\left[\pm(-k\eta)^2\mp\nu^2\right]^{3/2}} \left[2\nu^2+3(-k\eta)^2\right],\\
\left.\begin{array}{cc}
\omega_4\\
i\omega_4&
\end{array}\right\}&=\frac{1}{5760\left[\pm(-k\eta)^2\mp\nu^2\right]^{9/2}}\left[375(-k\eta)^6+3654\nu^2(-k\eta)^4+1512\nu^4(-k\eta)^2-16\nu^6\right],\\
\left.\begin{array}{cc}
\omega_6\\
i\omega_6&
\end{array}\right\}&=\mp\frac{1}{322560\left[\pm(-k\eta)^2\mp\nu^2\right]^{15/2}}\left[67599(-k\eta)^{10}+1914210\nu^2(-k\eta)^8+\right.\\\nonumber
&\left.+4744640\nu^4(-k\eta)^6+1891200\nu^6(-k\eta)^4+78720\nu^8(-k\eta)^2+256\nu^{10}\right],\\
\label{omega_8} \left.\begin{array}{cc}
\omega_8\\
i\omega_8&
\end{array}\right\}&=\frac{1}{3440640\left[\pm(-k\eta)^2\mp\nu^2\right]^{21/2}}\left[5635995(-k\eta)^{14}+318291750\nu^2(-k\eta)^{12}+\right.\\\nonumber &+1965889800\nu^4(-k\eta)^{10}+2884531440\nu^6(-k\eta)^8+1135145088\nu^8(-k\eta)^6+\\\nonumber
&\left.+99783936\nu^{10}(-k\eta)^4+881664\nu^{12}(-k\eta)^2-2048\nu^{14}\right],
\end{align}
where the upper and lower expressions on the left side and the upper
and lower signs on the right side in (\ref{omega_2})-(\ref{omega_8})
correspond to $-k\eta<-\nu$ and $-\nu<-k\eta<0$ respectively. After
computing $\omega(z)$, using the relations  (\ref{uk_right}) and
(\ref{uk_left}) we  obtain a ninth-order phase integral
approximation to the solution of the equation for scalar and tensor
perturbations (\ref{uk}). The constants $c_1$ and $c_2$ are computed
using the asymptotic behavior of $u_k$. We calculate the limit
$-k\eta\rightarrow \infty$ of the solution (\ref{uk}) on the left of
the turning point (\ref{uk_left}). Choosing $c_2=-ic_1$ with
$c_1=e^{i\left(\nu+\frac{1}{2}\right)\frac{\pi}{2}}/\sqrt{2}$ and
$c_2=e^{i\left(\nu-\frac{1}{2}\right)\frac{\pi}{2}}/\sqrt{2}$, we
obtain the asymptotic boundary condition  given by Eq.
(\ref{boundaryi}). In order to compute the power spectrum we need to
evaluate the limit $-k\eta\rightarrow 0$ for the growing part of the
solution (\ref{uk_right}). In this limit we have

\begin{equation}\label{uk_limit}
u_k^{phi}(\eta)\rightarrow\exp\left[i\left(\nu-\frac{1}{2}\right)\frac{\pi}{2}\right]f^{pi}_\nu
\frac{1}{\sqrt{k}} \left(-k\eta\right)^{\frac{1}{2}-\nu},
\end{equation}
where

\begin{equation}\label{fpi}
f^{phi}_\nu= \left(2\nu\right)^{\nu-\frac{1}{2}}
\exp\left(-\nu+\frac{1}{12 \nu}-\frac{1}{360 \nu^3}+\frac{1}{1260
\nu^5}-\frac{1}{1680 \nu^7}\right).
\end{equation}
Using  Eq. (\ref{uk_limit}), we have that the scalar and tensor spectra,
given by Eq. (\ref{PS_pi}) and Eq. (\ref{PT_pi}) are

\begin{eqnarray}
P_S^{phi}(k)&=&\frac{1}{l_0^2 M_{Pl}^2} \mathrm{g}^{phi}_\nu k^{3-2\nu},\\
P_T^{phi}(k)&=&\frac{1}{l_0^2} h^{phi}_\nu k^{3-2\nu},
\end{eqnarray}
where

\begin{eqnarray}
g^{phi}_\nu&=& \left(\frac{1-2\nu}{3-2\nu}\right)\left[\frac{f^{phi}_\nu}{2\pi}\right]^2,\\
h^{phi}_\nu&=&\left[\frac{2^{1/2}f^{phi}_\nu}{2\pi}\right]^2.
\end{eqnarray}
The spectral indices, given by Eq. (\ref{nS_pi}) and Eq. (\ref{nT_pi}) coincide with the
exact ones given in Eq. (\ref{indices1}) and Eq. (\ref{indices2}) respectively.
If we only keep the first term, $-\nu$, in the
exponential $f^{phi}_{\nu}$ (\ref{fpi}), we obtain the first-order
phase-integral approximation which coincides with the WKB method
after using  the Langer modification
\cite{langer:1937,martin:2003A,casadio:2005A}. If we keep the two
first terms in the exponential (\ref{fpi}), we obtain the
third-order phase integral approximation. It is worth mentioning
that, for the power-law model,  the tensor and scalar spectral
indices do not depend on the order of approximation.
\section{Results}\label{results}

\begin{figure}[htbp]
\begin{center}
\includegraphics[scale=0.50]{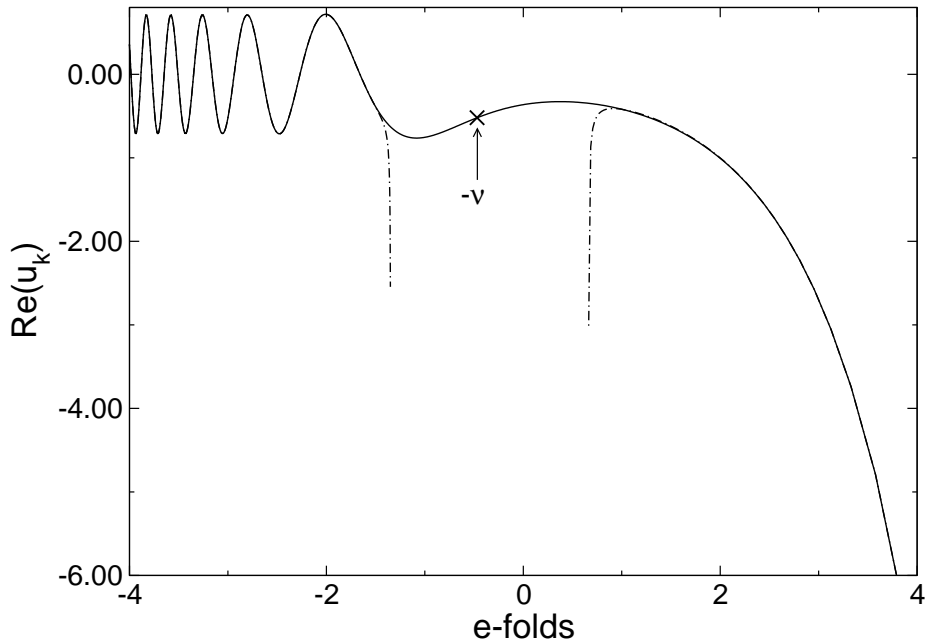}\\
\caption{$\Re(u_k)$ for the power-law inflationary model with $p=10$
with $k=1.389 h \mathrm{Mpc}^{-1}$. Solid line: analytic solution;
dot-dashed line: ninth-order phase integral approximation.}
\label{Re(uk)_pi9}
\end{center}
\end{figure}

In this section we proceed to compare the analytic solutions for  the wave function $u_k$ and the scalar and power
spectra   with the results obtained using the phase integral approximation.  Fig. \ref{Re(uk)_pi9} compares the real part
of the analytic solution of $u_k$  with the ninth-order phase
integral approximation. The plot is  made  using  the number of e-folds,
$N=\ln\ a$,  as the independent variable. As 
expected, the phase integral approximation solution diverges at the
root of $q(\eta)$

\vspace{2cm}

Now, we compare the scalar and tensor power
spectra calculated using the phase-integral approximation with the
results obtained with the slow-roll and uniform approximation
methods. From Ref. \cite{habib:2005B} (Eq. (63) and Eq. (64)) we
obtain that the scalar and tensor power spectra in the slow-roll
approximation are
\begin{eqnarray}
P_S^{sr}(k)=\frac{1}{l_0^2 M_{Pl}^2} g^{sr}_\nu k^{3-2\nu},\\
P_T^{sr}(k)=\frac{1}{l_0^2} h^{sr}_\nu k^{3-2\nu},
\end{eqnarray}
with

\begin{eqnarray}
g^{sr}_\nu= \left[1+2(2-\ln 2+b)(2\epsilon+\delta)-2\epsilon\right]\left(\frac{1-2\nu}{3-2\nu}\right)\left[\frac{2^{-\nu}\left| 1-2\nu\right|^{\nu-1/2}}{2\pi}\right]^2,\\
h^{sr}_\nu=\left[1-2(\ln
2+b-1)\epsilon\right]\left[\frac{2^{1/2-\nu}\left|1-2\nu\right|^{\nu-1/2}}{2\pi}\right]^2,
\end{eqnarray}
where b is the  Euler constant, $2-\ln2-b\simeq 0.7296$,
$\ln2+b-1\simeq 0.2704$, and $\epsilon=-\delta=\frac{1}{p}$. In the
slow-roll approximation $\epsilon \ll 1$, therefore, for the
power-law model, the slow-roll approximation is better suited for
large values of the parameter $p$.

Using the result obtained in Ref. \cite{habib:2004}  (Eq. (109)), we
obtain an expression for the second-order uniform approximation for
the power spectrum associated with the scalar and tensor
perturbations. They have the form:

\begin{eqnarray}
P_S^{ua}(k)=\frac{1}{l_0^2 M_{Pl}^2} g^{ua}_\nu k^{3-2\nu},\\
P_T^{ua}(k)=\frac{1}{l_0^2}h^{ua}_\nu k^{3-2\nu},
\end{eqnarray}
with

\begin{eqnarray}
\label{gua} g^{ua}_\nu=
\left(1+\frac{1}{6\nu}\right)\left(\frac{1-2\nu}{3-2\nu}\right)\left[\frac{(2\nu)^{\nu-1/2}e^{-\nu}}{2\pi}\right]^2,
\\ \nonumber
h^{ua}_\nu=\left(1+\frac{1}{6\nu}\right)\left[\frac{2^{1/2}(2\nu)^{\nu-1/2}
e^{-\nu}}{2\pi}\right]^2.
\end{eqnarray}
Omitting the
factor $1/{6\nu}$ in Eq. (\ref{gua})  we obtain the first-order uniform
approximation, result that coincides with the first-order
phase-integral approximation and the WKB method with the Langer
modification \cite{martin:2003A}. Keeping the second term of the
expressions in Eq. (\ref{gua}) one gets the second-order uniform
approximation.

\begin{figure}[htbp]
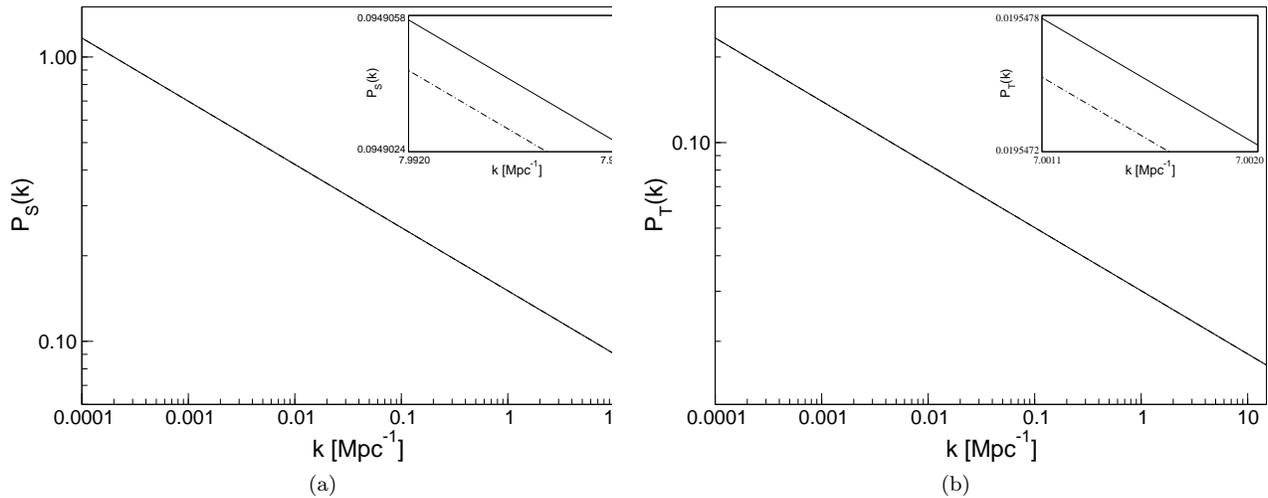

\begin{center}
\subfigure[]{\includegraphics[scale=0.35]{pl_doublePS_ex,pi9.eps}}
\subfigure[]{\includegraphics[scale=0.35]{pl_doublePT_ex,pi9.eps}}\\
\caption{(a) Scalar power spectrum  $P_S(k)$ and (b) tensor power
spectrum  $P_T(k)$ for the power-law inflationary model with $p=10$.
The solid line indicates the analytic solution.  The dot-dashed line:
ninth-order phase integral approximation. The inset in each figure represents
an enlargement that permits one to compare the accuracy of each method.} \label{PST1}
\end{center}
\end{figure}

\begin{figure}[htbp]
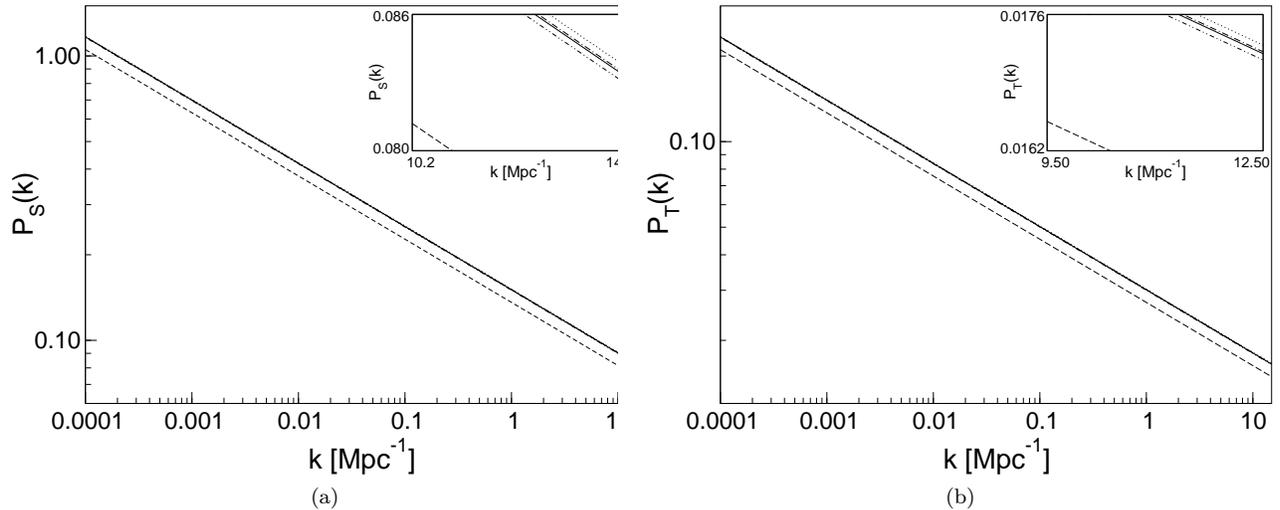

\begin{center}
\subfigure[]{\includegraphics[scale=0.35]{pl_doublePS_ex,pi3,ua1,ua2,sr.eps}}
\subfigure[]{\includegraphics[scale=0.35]{pl_doublePT_ex,pi3,ua1,ua2,sr.eps}}\\
\caption{(a) Scalar power spectrum  $P_S(k)$ and (b) tensor power
spectrum $P_T(k)$  for the power-law inflationary model with $p=10$.
Solid line: Analytic solution; dotted line: slow-roll approximation;
dashed line: first-order phase-integral approximation, WKB and
first-order uniform approximation; dot-dashed line: third-order
phase integral approximation; dot-dot-dashed line: second-order
uniform approximation. The inset in each figure represents
an enlargement that permits one to compare the accuracy of each method.}
\label{PST3}
\end{center}
\end{figure}

We want to compare the analytic expression for the scalar and tensor
power spectra for different values of $k$  with  the ninth-order phase integral
 approximation, the
slow-roll approximation and the first and second-order uniform
approximation. Fig. \ref{PST1} and Fig. \ref{PST3}  show the power
spectra $P_S(k)$ and $P_T(k)$ calculated analytically as well as with
different approximation methods. The inset in each figure represents
an enlargement that permits one to compare the accuracy of the different methods.
We stop the computation of $P_S(k)$ and
$P_T(k)$ when the quotient $u_k/z$ (scalar perturbations) or $u_k/a$
(tensor perturbations) becomes constant, i.e., when the function
$u_k$ leaves the horizon.

\begin{figure}[htbp]
\begin{center}
\includegraphics[scale=0.50]{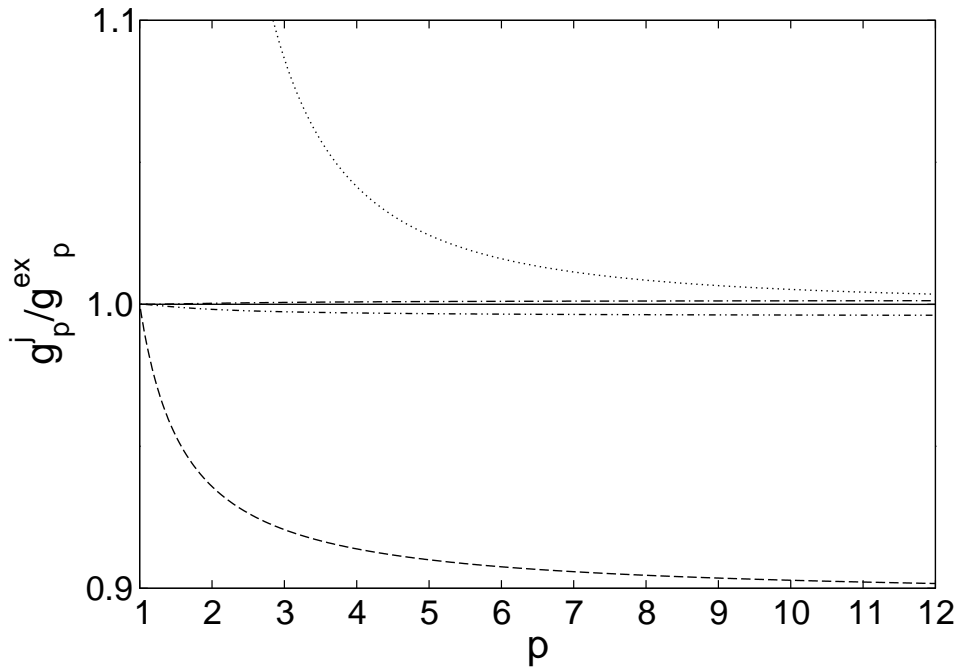}\\
\caption{Evolution of the ratio $g^j(p)/g^{ex}(p)$ versus
$p$. Solid line: ninth-order phase integral approximation; dotted
line: slow-roll approximation; dashed line: first-order phase
integral, WKB and first-order uniform approximations; dot-dashed
line: third-order phase-integral approximation; dot-dot-dashed
line: second-order uniform approximation.} \label{coeficient}
\end{center}
\end{figure}

Fig. \ref{coeficient} shows the quotient between the exact analytic
result and the result obtained using different methods of
approximation  $g^\ji_p/g^\ex_p$ for different values of $p$. Using
the ninth-order phase integral approximation we obtain a horizontal
line equal to unity. We observe analogous behavior when we
consider the quotient $h^\ji_p/h^\ex_p$. As it was already mentioned,
the slow-roll approximation is better suited for large values of $p$
because $\epsilon=1/p$. The WKB method gives a better approximation
for small values of  $p$ because the condition $\mu \ll 1$ is valid
in this case. It should be noticed that the first-order
phase-integral approximation, the WKB method with the Langer
modification and the first-order uniform approximation give the same
result.

\begin{figure}[htbp]
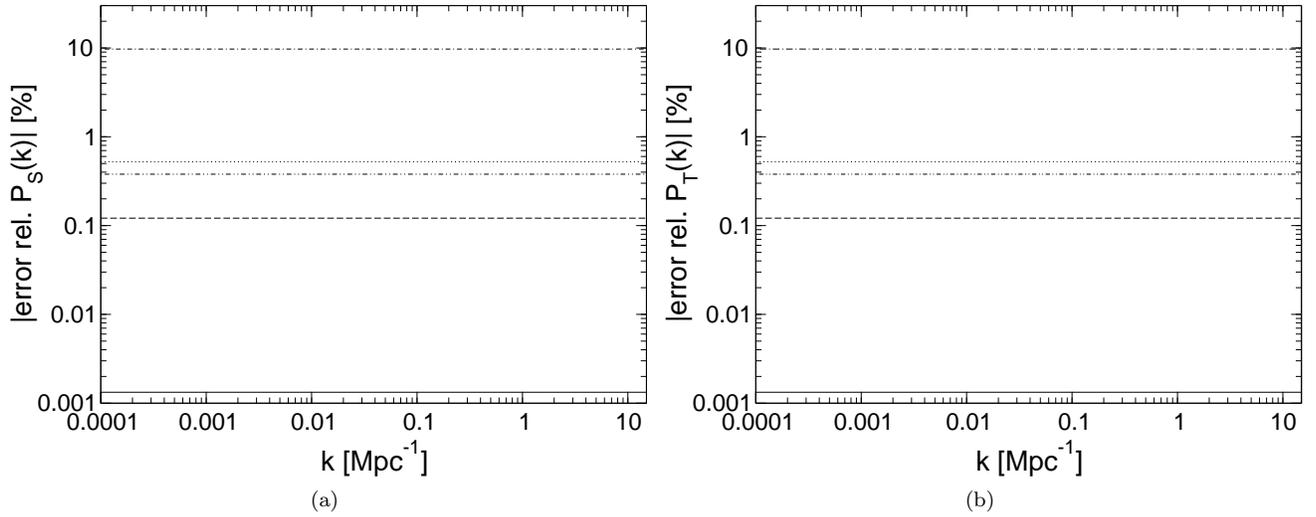

\begin{center}
\subfigure[]{ \label{pl:error_PS}
\includegraphics[scale=0.35]{pl_error_PS_pi3,pi9,wkb,ua1,ua2,sr.eps}}
\subfigure[]{ \label{pl:error_PT}
\includegraphics[scale=0.35]{pl_error_PT_pi3,pi9,wkb,ua1,ua2,sr.eps}}
\caption{(a) Relative error of  $P_\sca(k)$ and (b) relative error
of $P_\ten(k)$  for the power-law inflationary model with $p=10$.
Solid line: ninth-order phase integral approximation, dashed line:
third-order phase integral approximation; dot-dot-dashed line:
second-order uniform approximation; dotted line : slow roll
approximation; dot-dashed line: first-order phase integral
approximation, $\text{WKB}$, and first-order uniform
approximation}\end{center}
\end{figure}

Fig.\ref{pl:error_PS} and Fig.\ref{pl:error_PT} show,
the relative error of $P_\sca(k)$ and  $P_\ten(k)$ respectively for different
approximation methods. The relative error is computed using the
following expression:
\begin{equation}
\text{rel. error} \,\,P_{\sca,\ten}(k)
=\frac{\left[P_{\sca,\ten}^\text{\text{approx}}(k)-P_{\sca,\ten}^\ex(k)\right]}{P_{\sca,\ten}^\ex(k)}\times
100.
\end{equation}

\begin{table}[htbp]
\begin{center}
\begin{tabular}{c|c|c}
\hline\hline
Approximation &$\left|\text{rel. error}\,\,  P_\sca(k)\right|$ & $\left|\text{rel. error}\,\, P_\ten(k)\right|$\\
\hline
phi9$^\text{\tiny{a}}$  &  $ 0.0015 \% $        & $ 0.0015\%$     \\
phi3$^\text{\tiny{b}}$   &  $ 0.1\% $        & $ 0.1\%$     \\
ua2$^\text{\tiny{c}}$ &  $ 0.4\% $        & $0.4\%$     \\
sr$^\text{\tiny{d}}$                      &  $  0.5\% $        & $0.5\%$     \\
phi1$^\text{\tiny{e}}$, $\textnormal{WKB}_\mathrm{LM}^\text{\tiny{f}}$, ua1$^\text{\tiny{g}}$ &    $10\% $        & $10\%$ \\
\hline\hline
\end{tabular}
\end{center}
\caption[Relative error obtained using different methods of
approximation for the power-law inflationary model with $p=10$ and
$k=1.369\,h\,\Mpc^{-1}$]{\small{Relative error obtained using
different methods of approximation for the power-law inflationary
model with $p=10$ for the mode $k=1.369\,h\,\Mpc^{-1}$. }}
\label{pl_t2}
\begin{flushleft}
$^\text{\tiny{a}}$Ninth-order phase integral approximation.\\
$^\text{\tiny{b}}$Third-order phase integral approximation.\\
$^\text{\tiny{c}}$Second-order uniform approximation.\\
$^\text{\tiny{d}}$Slow-roll approximation.\\
$^\text{\tiny{e}}$First-order phase integral approximation.\\
$^\text{\tiny{f}}$ $\textnormal{WKB}$ approximation with the Langer modification.\\
$^\text{\tiny{g}}$First-order uniform approximation.\\
\end{flushleft}
\end{table}

Table \ref{pl_t2} shows that the ninth-order phase integral
approximation gives the smallest relative error for $P_\sca(k)$ and
$P_\ten(k)$. The third-order phase integral approximation gives a
relative error smaller than the one obtained using the second-order
uniform approximation.

Finally, some words about the accuracy of the phase-integral
approximation are in order: The smallness of the integral
$\mu(z,z_{0})$ given by Eq. (\ref{mu})  is a measure of the accuracy
of the phase-integral approximation. Fig.  \ref{Re(uk)_pi9} shows
that the phase-integral method fails in the vicinity of the turning
point $-\nu$, range where the $\mu$-integral diverges. The selection
of the base function $Q(z)$ given by Eq. (\ref{Q2}) guarantees
that $\mu\ll 1$  far from the turning point  at any order of
approximation.  Since the scalar and tensor power spectra as well as
the spectral indices are evaluated as $-k\eta\rightarrow 0$, the
limit  is taken far from the horizon (turning point), therefore
their computation is not affected by the presence of the turning
point.  For  $p=10$ and  $k=1.369 h Mpc^{-1}$ the turning point is
$\nu=-1.61 M_{Pl}^{-1}$, and the power spectrum is evaluated at
$k\eta=-10^{-8} M_{Pl}^{-1}$, a point  in the limit where the
function $u_k$ exhibits the asymptotic behavior (\ref{boundary_0}).
Fig.\ref{PST1}, Fig. \ref{PST3} and Table \ref{pl_t2} show the
accuracy of the phase-integral approximation.

In the present paper we have shown that, in comparison with other
approximation methods, the phase integral approach gives very good
results  for the scalar and tensor spectra in the power-law
inflationary model. We have also seen that the  phase integral approach
reproduces the exact spectral indices in the power-law model. Since
the WKB method can be regarded as a first-order approximation of the
phase-integral approximation with $Q^2(z)=R(z)$, it should be
expected that the phase-integral method works in those cases where
the WKB methods gives good estimates and  slow-roll fails, that is the
case where inflation is generated by a chaotic potential with a step
\cite{Hunt}.  The good agreement between the analytic results and
the phase-integral approximation shows that the
phase integral method is a very useful tool for
computing the scalar and tensor  power spectra in more realistic
inflationary scenarios, in a forthcoming work we will
implement the phase integral method to quadratic and quartic chaotic
models.

\acknowledgments

We thank Dr. Ernesto Medina for reading and improving the
manuscript. This work was partially supported by FONACIT under
project G-2001000712.


\end{document}